\documentclass{Interspeech2024}

 \interspeechcameraready

\title{Learnings from curating a trustworthy, well-annotated, and useful dataset of disordered English speech}

\name[affiliation={1}]{Pan-Pan}{Jiang}
\name[affiliation={1}]{Jimmy}{Tobin}
\name[affiliation={1}]{Katrin}{Tomanek}
\name[affiliation={1}]{Robert L.}{MacDonald}
\name[affiliation={2}]{Katie}{Seaver}
\name[affiliation={3}]{Richard}{Cave}
\name[affiliation={4}]{Marilyn}{Ladewig}
\name[affiliation={1}]{Rus}{Heywood}
\name[affiliation={2}]{Jordan R.}{Green}

\address{
  $^1$Google Research
  $^2$MGH-IHP
  $^3$Language and Cognition UCL
  $^4$CPUnlimited}
\email{pjiang@google.com, jtobin@google.com}


\begin{document}

\maketitle

\begin{abstract}
    
    % 1000 characters. ASCII characters only. No citations. 1000 character limit
Project Euphonia, a Google initiative, is dedicated to improving automatic speech recognition (ASR) of disordered speech. A central objective of the project is to create a large, high-quality, and diverse speech corpus. This report describes the project's latest advancements in data collection and annotation methodologies, such as expanding speaker diversity in the database, adding human-reviewed transcript corrections and audio quality tags to 350K (of the 1.2M total) audio recordings, and amassing a comprehensive set of metadata (including more than 40 speech characteristic labels) for over 75\% of the speakers in the database. We report on the impact of transcript corrections on our machine-learning (ML) research, inter-rater variability of assessments of disordered speech patterns, and our rationale for gathering speech metadata. We also consider the limitations of using automated off-the-shelf annotation methods for assessing disordered speech.
\end{abstract}

\section{Introduction}
\label{sec:intro}

 The Project Euphonia \cite{macdonald2021euphonia} data corpus is, to our knowledge, the largest dataset of disordered English speech in the world, and currently comprises over 1.2M million utterances recorded from $\sim$2000 speakers with varied etiologies. The long-term goal of the project is to make speech recognition technology accessible to everyone, regardless of their speech abilities, similar to work in \cite{tobin2022personalized,Lea2023Stuttering,green2021automatic}. This paper outlines the progress and scale of Project Euphonia's dataset, emphasizing its role in improving speech recognition for people with speech disorders. It details enhancements in data diversity, trustworthiness, and collection efficiency, aiming to support the development of a high-quality speech corpus for accelerating ML-driven research on disordered speech.

\section{Expanding Data Diversity}
\label{sec:dataset_imp}
\subsection{Increasing the diversity of speakers and etiologies }
\label{subsec:prompted}

We increased our effort to enroll speakers from a broad spectrum of backgrounds, including individuals from historically underrepresented racial and socioeconomic groups and covering a wide variety of etiologies of speech disorders. Our approach involved forming strategic alliances with diverse organizations dedicated to serving people with communication impairments such as ADAPT Community Network and LSVT Global.  In total, our corpus includes over 2000 speakers with varied etiologies including amyotrophic lateral sclerosis (ALS), Parkinson’s disease (PD), Down syndrome (DS), laryngectomy (LAR), vocal cord paralysis (VCP), hearing impairment (HI), multiple sclerosis (MS), cerebral palsy (CP). In addition, to make our data collection efforts more inclusive, we tailored our strategy to cater to participants with different literacy levels. This included the creation of a new set of prompts, composed of shorter phrases and utilizing vocabulary suitable for a Grade 3 to Grade 5 reading level, as measured by the Flesch-Kincaid scale \cite{flesch1948new}.

\subsection{Documenting diversity of atypical speech patterns}
\label{subsec:diversity}
To ascertain the diversity of the types of atypical speech patterns in our dataset, we developed a human-centric labeling system that could effectively capture a wide range of speech and voice irregularities. This method involved the use of evaluations by certified speech-language pathologists (SLPs), who received specialized training in assessing and grading speech abnormalities. Given the substantial heterogeneity of speech impairments, we graded each sample based on 40 different disordered speech labels. The grading scheme was consistent across the 40 different labels, with most rated along a five-point equal interval scale: typical (no impairment), mild impairment, moderate impairment, severe impairment, and profound impairment. These labels were generated for each of the speech subsystems: articulatory, phonatory, resonatory, and respiratory, and for global measures of speech such as overall severity, speech intelligibility, speaking rate, speech pattern consistency, and prosody. Additional labels were evaluated for the phonatory system including vocal pitch and loudness, and dysphonia characteristics, which included roughness, breathiness, strain, and tremor. 

While tools such as openSMILE \cite{eyben2010opensmile} provide options for automatically labeling disordered speech and voice, our choice to employ expert human raters was driven by the ongoing debate over the clinical validity of quantitative acoustic metrics \cite{stegmann2020repeatability,vogel2011reliability}. The substantial operational costs of obtaining expert human labels, however, necessitated the development of a brief yet comprehensive labeling system, enabling the utilization of a single rater instead of relying on multiple raters or reaching a consensus. In total, we amassed a comprehensive set of these annotations for over 75\% of the speakers in our database.

\subsection{Expanding linguistic and speech pattern diversity}
\label{subsec:prompted}
To enhance the linguistic diversity of our dataset, we transitioned from having all participants record identical phrases to a system where each participant recorded a few phrases randomly selected from a broader pool. We also extended our data collection efforts to include spontaneous speech samples. Unlike the more controlled, prompted speech, spontaneous speech is produced with less precision \cite{kuo2016vowel} and may be less intelligible 
\cite{kempler2002effect, weir2017internally}.  To collect spontaneous speech samples, we initiated a Trusted Tester Program, inviting a select group of participants who were interested in contributing to our research and willing to provide additional consent. These datasets consisted of both speech elicited using assessor-prompted conversation cues and fully self-generated speech. Before integrating these samples into our research corpus, we conducted a thorough review to identify and remove any Personally Identifiable Information (PII), ensuring the privacy and security of the data. The samples were added to our research corpus only after this careful screening process.

\section{Improving trustworthiness of corpus}
\label{sec:human_annotation}
\subsection{Manual data validation and cleaning}
\label{subsec:trustworthiness}
As our corpus of disordered speech grew in size, human intervention was required to facilitate the detection and removal of low-quality data, such as recordings with poor audio quality, empty recordings, and recordings made by a bot or spammer. We also standardized transcriptions to normalize all prompts to follow the same transcription guidelines (e.g., “6 am” vs “6:00 a.m.”). Lastly, we created human-reviewed and corrected transcripts for 352,130 utterances, which accounts for 29\% of our overall corpus, with a focus on reviewing test utterances ($>40\%$ reviewed at the time of writing).

We ran experiments to understand the benefit of transcript correction on model accuracy. We chose to focus our assessments on speakers with Down Syndrome since these speakers tended to require more transcript corrections (usually due to reading errors), thus affording us an opportunity to assess any effects those corrections may have. In the end, from a pool of 1000 speakers, we were able to find nine speakers with Down Syndrome (all rated as having an overall mild severity of speech impairment) for whom we had an adequate number of recordings (at least 350), and who had at least 15\% of their transcripts reviewed, which in this case resulted in between 700 and 3800 manually corrected transcripts. 

On average we standardized 72\% of each speaker’s transcripts, with the vast majority of corrections (95\% on average) due to punctuation, capitalization, and other normalization aspects of the transcript. Actual word changes (i.e., misreadings) occurred on $< 5\%$ of the utterances on average. While many of these transcription errors may not have been meaning-changing, they did impact reported WERs.

We analyzed the impact of corrected transcripts on the training process. We trained models in a similar fashion to \cite{tobin2022personalized}, with and without corrected transcripts. After evaluating on corrected test transcripts, results from training with corrected transcripts were inconclusive. For some speakers, we saw improvements; for others, we saw small deteriorations (overall, there was no statistically significant improvement across the 9 speakers).

We then analyzed the impact of corrected transcripts on model quality assessment (evaluation). WERs decreased when using corrected transcripts (compared to using uncorrected transcripts) in 7 of the 9 speakers (with deltas ranging from -0.2 WER to -4.5 WER). And increased in 2 of the 9 speakers (deltas of +0.3 WER and +0.9 WER). This meant that most of the time, assessing WERs on uncorrected transcripts led to an overestimation of WERs.

Based on these results, we concluded that transcript normalizations should be consistently applied across all transcripts. Consequently, for Project Euphonia, we automated the application of transcript normalizations for our own data collection operations. 
 
\subsection{Testing automated approaches for identifying low-quality data
}
\label{subsec:metadata}
Manual quality control and labeling are expensive and subjective. Therefore, our group has been exploring a variety of methods to automate quality control. To date, our efforts have primarily focused on developing tools that can automatically detect poor-quality audio recordings. Recording quality can vary from participant to participant and even from recording session to recording session. In some cases, recordings do not contain speech because of user error. Below, we outline our methodology for evaluating a Voice Activity Detection (VAD) strategy aimed at identifying recordings within our disordered speech dataset that, in error, did not contain speech.

VAD models classify a sequence of acoustic frames as either speech or silence \cite{sohn1999statistical}. A VAD is part of the process of endpointing, which can filter and segment speech recordings before the recordings are run through an ASR model. VAD precision is known to be affected by the level of noise in a speech recording, but \cite{ding20_odyssey} shows that precision for classifying typical speech reduces from 90\% to 80\% when noise is introduced into a signal. 
We analyzed the performance of two off-the-shelf VAD models (VAD1, a cloud-based speech VAD similar to \cite{chang2019endpointer}, and VAD2, which was designed for on-device ASR similar to \cite{sainath2020streaming})  for detecting voice in our recordings of disordered speech. 

\begin{figure}[t]
  \centering
  \includegraphics[width=\linewidth]{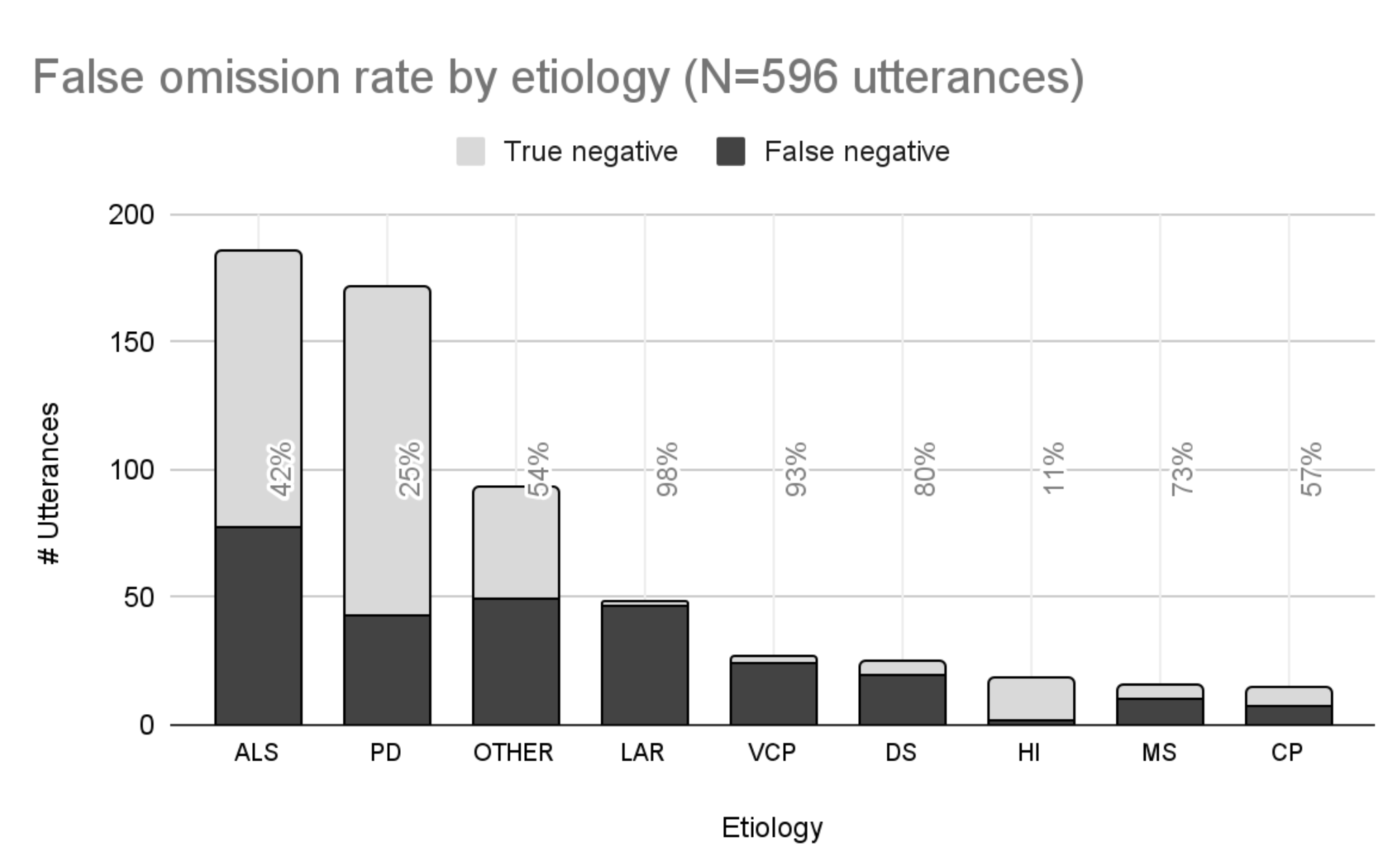}
  \includegraphics[width=\linewidth]{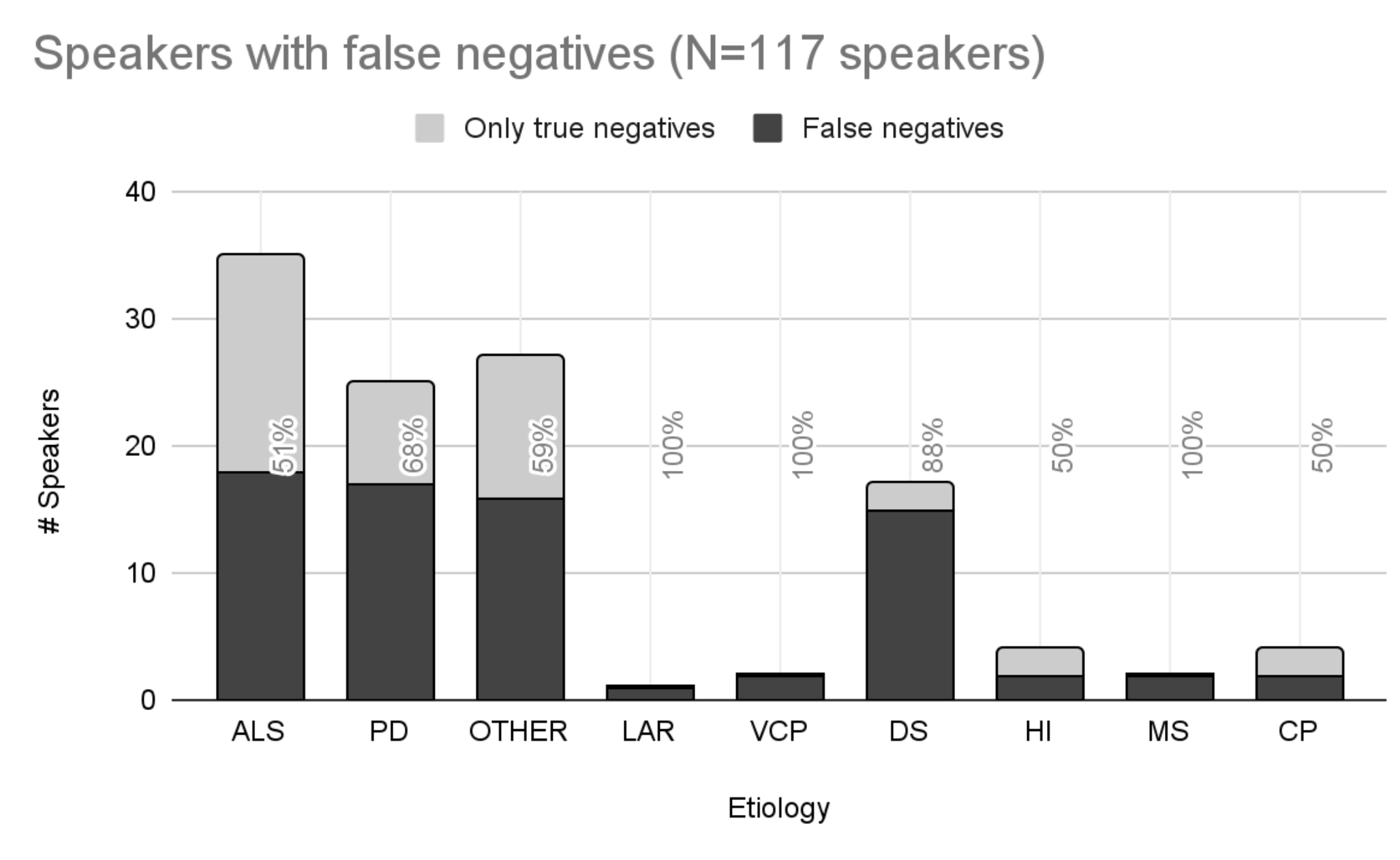}
  \caption{Analyzed VAD negative decisions grouped by etiology. False omission rate is labeled for each etiology. (Top) utterance counts, (bottom) speaker counts.}
  \label{fig:false_omission_rates}
\end{figure}

VAD1 was run on 825k utterances. $\sim$8k utterances were found to be marked as not containing speech. SLPs listened to 7.5\% of the $\sim$8k utterances (596 utterances) to determine the false omission rate (Figure~\ref{fig:false_omission_rates}). The false omission rate is defined as the number of utterances marked as not containing speech which actually contained speech. Of the 378 total speakers who had utterances without detected speech, 117 speakers (31\%) had utterances assessed (Figure~\ref{fig:false_omission_rates}). We found that VAD1 had a false omission rate of 47.7\%. This false omission rate is evidence that the atypical speech patterns have an effect distinct from signal noise alone where we would expect 80\% precision. Further, certain etiologies were associated with the highest false omission rates including LAR, VCP, MS, and DS. 

Speech systems that employ VADs may unintentionally worsen the user experience for individuals with speech disabilities. Specifically, within numerous speech applications, the VAD might exclude certain audio samples from transcription.

To simulate the effect of using a VAD to automatically reject speech recordings from a disordered speech corpus, we selected a subset of 20 speakers with etiologies mentioned above as being most affected by VAD bias and compared the performance of VAD1 to VAD2 on their 32,000 speech recordings. Based on these simulations, we found that a pipeline with VAD1 would reject $\sim$4,000 recordings, while one with VAD2 would reject $\sim$500 recordings. Therefore, the use of VAD1 instead of VAD2 would have resulted in the rejection of on average 175 utterances for each speaker (which is equivalent to an hour or more of user effort). This difference is evidence that not all VAD models are created equal and care should be taken when they are integrated into data curation pipelines. A more permissive VAD is one option, but a VAD trained with impaired speech is very likely a better solution. As a result, we do not use VAD models in any part of our data curation pipeline.

\subsection{Establishing the replicability of human expert labeling }
\label{sec:rater_rel}
\begin{figure}[t]
  \centering
  \includegraphics[width=\linewidth]{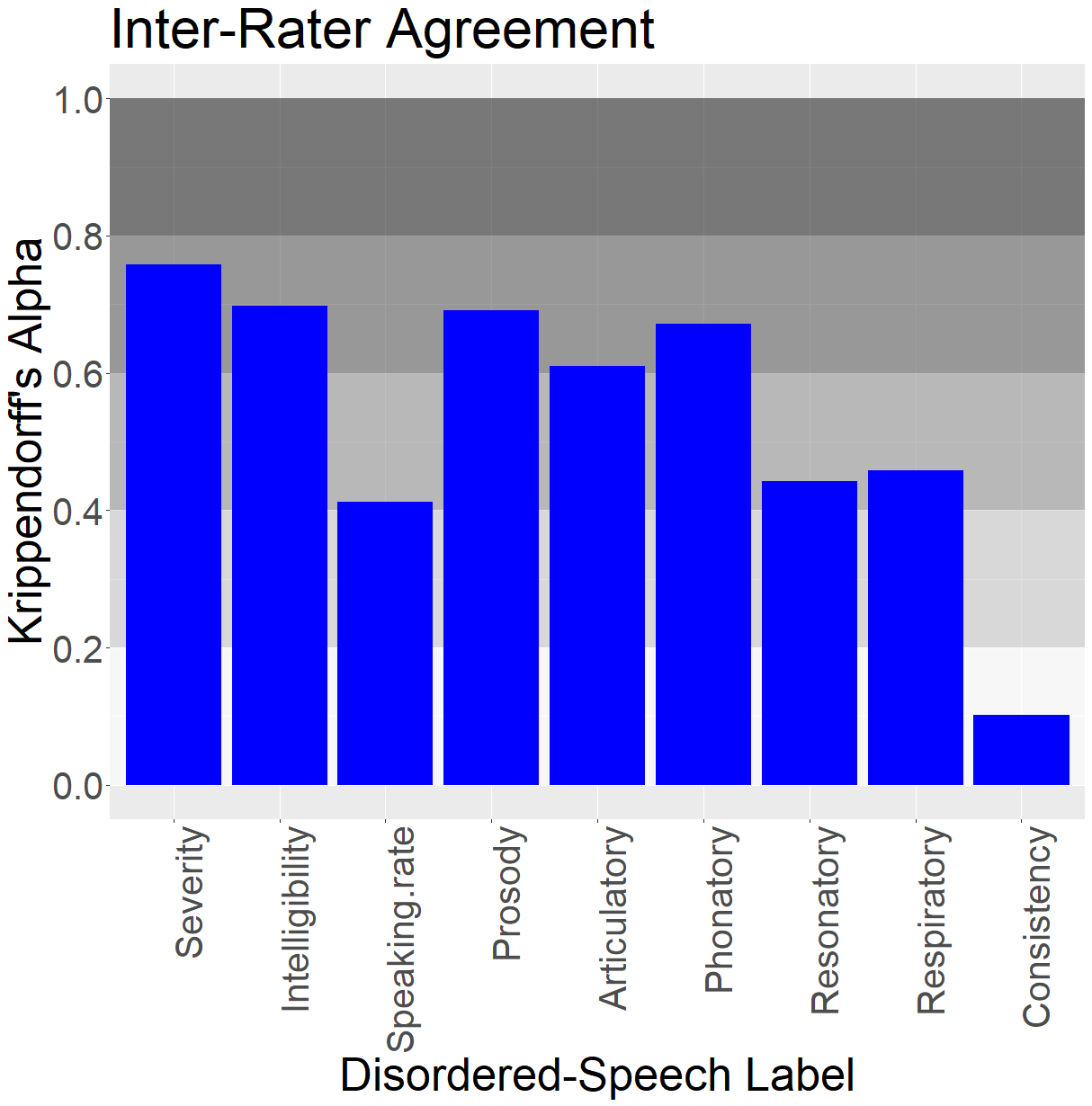}
  \caption{Estimates of inter-rater reliability for nine of the disordered speech labels.}
  \label{fig:inter-rater_rel}
\end{figure}

Reliable and accurate labeling is essential in ML research, providing the foundation for developing models that are effective and trustworthy. We tested the inter-rater reliability of our disordered speech labels by having two SLPs independently rate 50 recordings (Figure~\ref{fig:inter-rater_rel}). Krippendorff’s alpha \cite{hughes2021krippendorffsalpha} showed that reliability can vary significantly across labels. Although most of the labels achieved substantial or moderate agreement, the consistency and speaking rate labels were unreliable. In contrast, the severity label was highly reliable, a finding that is consistent with prior reports \cite{stipancic2021you}. Taken together, these findings support the need to establish the reliability of disordered speech labels prior to deploying them for the purpose of model building.

\section{Improving Data Collection Efficiency}
\label{sec:efficiency}
Prior research has shown that even small amounts of speech recordings (20 minutes) can be used to personalize on-device automatic speech recognition (ASR) models for use in home automation and caregiver situations \cite{tobin2022personalized}. Based on this finding, we reduced the number of phrases assigned per user from 1500 to 300. To further improve data collection efficiency, we curated our prompt lists to eliminate prompts that were inappropriate for our users (e.g., too much emphasis on physical mobility) or too difficult to pronounce. We also better-aligned prompts with use cases of interest for our target populations (e.g., home automation, voice search queries).

\section{Data Access}
\label{sec:access}

The absence of large, well-curated speech-disordered databases has posed a substantial hurdle to advancements in our field. While the primary aim of this paper is to detail the processes, procedures, and underlying rationale for curating this novel dataset, we understand researchers will be most interested in the dataset itself, the value of which transcends academic and industry boundaries. As such, we helped establish  the Speech Accessibility Project (SAP - https://speechaccessibilityproject.beckman.illinois.edu) - a collaboration between researchers at the University of Illinois Urbana-Champaign (UIUC), and five technology companies, including Google. The SAP aims to collect and curate datasets of impaired speech that will be made available to requestors who sign the UIUC's Data Use Agreement and whose application is deemed aligned with the program's objectives by UIUC. Project Euphonia regularly provides advisory support to the SAP based in large part on learnings represented in this paper. Most importantly: Project Euphonia’s ~2000 participants, whose contributions comprise the 1.2M utterance dataset herein described, can choose to share their recordings with the SAP so that their data can be studied more broadly.

\section{Summary and Future Directions}
\label{sec:summary}

In this report, we described the iterative improvements made to our dataset of disordered English speech, including updates to the speech sets participants are asked to record; manual edits of recorded audio to create groundtruth transcripts; an adoption of annotation systems developed to create useful metadata and labels; and research into the limitations of some off-the-shelf automated speech label technologies for disordered speech. The following improvements to our annotation processes and datasets are also now being considered by the team:

\begin{itemize}
    \item Exploring the efficacy of potentially more granular and reliable speech labeling approaches, such as visual analog scales (VAS), direct magnitude scaling (DMS), or consensus-based ratings. \cite{schiavetti1992scaling}. 
    \item Improving automated annotation technologies, which have to date been trained only on typical speech samples, by incorporating disordered speech datasets. 
    \item Expanding our models to include Indic, Asian, and European languages.
\end{itemize}

These improvements will enhance ASR by providing more accurate and diverse labeling, training the technology with a broader range of speech patterns, and expanding linguistic coverage, ensuring a more inclusive and precise recognition across different languages and speech disorders.

\bibliographystyle{IEEEtran}
\bibliography{mybib}

\end{document}